# Fluorescence-Guided Raman Spectroscopy for Tumour Margin Delineation

Conor C. Horgan[1,2,3], Mads S. Bergholt[1,2,3#], May Zaw Thin[4], Anika Nagelkerke[1,2,3†], Robert Kennedy[5], Tammy L. Kalber[4], Daniel J. Stuckey[4], *Molly M. Stevens[1,2,3]

[1]Department of Materials, Imperial College London, London SW7 2AZ, UK.
[2]Department of Bioengineering, Imperial College London, London SW7 2AZ, UK.
[3]Institute of Biomedical Engineering, Imperial College London, London SW7 2AZ, UK.
[4]Centre for Advanced Biomedical Imaging, University College London, London WC1E 6DD, UK.
[5]Guy's and St Thomas' NHS Foundation Trust, Oral/Head and Neck Pathology, King's College London, London SE19RT, UK.
[#]Current address: Centre for Craniofacial and Regenerative Biology, King's College London, London SE1 9RT, UK.
[†]Current address: Groningen Research Institute of Pharmacy, Pharmaceutical Analysis, University of Groningen, Groningen NL-9700 AD, The Netherlands.
* Corresponding Author: m.stevens@imperial.ac.uk

## Abstract

The initial detection and identification of suspicious lesions and the precise delineation of tumour margins are essential for successful tumour resection, with progression-free survival linked to rates of complete resection. However, post-surgical positive margin rates remain high for many cancers and despite numerous advances in intraoperative imaging and diagnostic technologies, there exists no single modality that can adequately perform both tumoural detection and delineation. Here, we demonstrate a multimodal computer vision-based diagnostic system capable of both the gross detection and identification of suspicious lesions and the precise delineation of disease margins. We first show that through visual tracking of a spectroscopic probe, we enable real-time tumour margin delineation both for *ex vivo* human tumour biopsies and for an *in vivo* tumour xenograft mouse model. We then demonstrate that the combination of Raman spectroscopic diagnoses with protoporphyrin IX (PPIX) fluorescence imaging enables fluorescence-guided Raman spectroscopic margin delineation. Our fluorescence-guided Raman spectroscopic system achieves superior margin delineation accuracy to fluorescence imaging alone, demonstrating the potential for our system to achieve improved clinical outcomes for tumour resection surgeries.

## Introduction

Accurate delineation of tumour margins is essential for improving cancer survival rates as incomplete tumour resection has been shown to significantly reduce long-term survival rates for a range of cancers[1–3]. However, the need for maximal resection needs to be balanced with the goal of healthy tissue preservation in order to minimise patient discomfort and functional impairment. Many groups have thus been researching advanced imaging and spectroscopic techniques for improved tumour visualisation and margin delineation[4–7]. Fluorescence-guided surgery (FGS) has, for example, been employed with great success for the visualisation of high-grade gliomas using fluorophores, enabling improved rates of complete resection and progression-free survival relative to conventional microsurgery[8–10]. This technique has however been somewhat limited by difficulties in the quantification of fluorescence levels due to varying tissue optical properties and low sensitivities for early-stage cancers[11–13].

As an alternative, many groups have investigated the application of pointwise optical techniques such as fluorescence spectroscopy[7,14,15], reflectance spectroscopy[16–18], Raman spectroscopy[4,19,20], optical coherence tomography[21–23], and confocal endomicroscopy[24,25] for cancer detection and diagnosis. These techniques probe the optical or endogenous biomolecular properties of the tissue itself, revealing differences between healthy and diseased tissue that can be used to provide accurate diagnosis. For example, fluorescence spectroscopy has been applied to skin cancer diagnosis[26,27] and as a quantitative adjunct to fluorescence imaging of gliomas during brain surgery[11,15]. Reflectance spectroscopy has been used for the detection of cervical precancers *in vivo*[16] and in combination with fluorescence spectroscopy for the *in vivo* detection of breast, brain, and ovarian cancers[17,18,28]. Similarly, Raman spectroscopy has enabled highly accurate *in vivo* diagnosis of a range of cancers including breast, skin, colon, gastric, and oesophageal cancers, exploiting the interaction of light with molecular bonds to identify the chemical species present in a sample[4,19,29–31].

Though each of these spectroscopic modalities has shown promise in the accurate diagnosis of cancerous tissues, clinical utilisation is hindered by the lack of widefield imaging. As spectroscopic techniques, these methods typically involve the collection of point spectra via a handheld fibreoptic probe, providing diagnostic information at discrete locations, rather than a diagnostic image for a large area, as is the case with fluorescence imaging[18]. As such, although they can provide highly accurate diagnoses of a given point, tumour margin delineation is unrealistic unless clinicians can visualise and record diagnoses for multiple points simultaneously[32]. Further, particularly in the case of Raman spectroscopy, pixel-by-pixel measurement of a surgical field-of-view (FOV) is infeasible owing to infrequency of Raman scattering events, which necessitates relatively long acquisition times to generate sufficient signal[33]. Importantly, however, these diagnostic modalities offer different, yet complementary information. While widefield imaging provides crucial macroscopic morphological information, it lacks the microscopic biochemical information offered by many pointwise optical techniques. Therefore, the development of a system that effectively combines spatially co-registered spectroscopic diagnostic information with widefield imaging could vastly improve the clinical utility of handheld spectroscopic probes for cancer diagnostics.

Here we present a new approach for the acquisition of spatial spectroscopic diagnostic information via computer vision tracking of a handheld spectroscopic probe. Our system enables simultaneous recording of both the position and orientation (pose) of the spectroscopic probe as well as the diagnostic spectroscopic information for each measurement acquisition. Together, the data is overlaid onto imaging of the surgical FOV to provide a near real-time augmented reality (AR) display of the FOV. We demonstrate our system is capable of near real-time operation for accurate lesion mapping both *ex vivo* and *in vivo* under white light image guidance, providing comprehensive clinical control over diagnostic parameters to enable system tuning to varied clinical contexts. We further show that our system can be extended to include fluorescence image guidance, resulting in improved margin delineation of fluorescent optical tissue phantoms than fluorescence imaging alone. Our fluorescence-guided Raman spectroscopic system thus helps to bridge the gap between point-based spectroscopic diagnoses and imaging information, overcoming the trade-off between diagnostic accuracy and FOV that has to date limited spectroscopic diagnostic systems.

## Results

### Fluorescence-Guided Raman Spectroscopic System Development

Our fluorescence-guided Raman spectroscopic system combines widefield imaging with Raman spectroscopic information, and consists of a camera for white light/fluorescence imaging, a handheld fibreoptic probe, a laser and spectrograph for Raman spectroscopy, an excitation light source, collection filter optics for fluorescence imaging, and a computer with integrated software for clinical control (Figure 1). Clinical application of our system necessitates a probe-tracking algorithm that is robust under varied settings including different lighting conditions, imaging devices, and spectroscopic probes. We therefore aimed to develop a system that could be readily translated across different environments with minimal technical requirements while maintaining real-time capabilities. To achieve this, we implemented a marker-based visual tracking algorithm that combines visual detection and tracking of coloured markers with *a priori* knowledge of probe geometry to determine the position and orientation (pose) of a probe for spatial diagnostics (Figure 2).

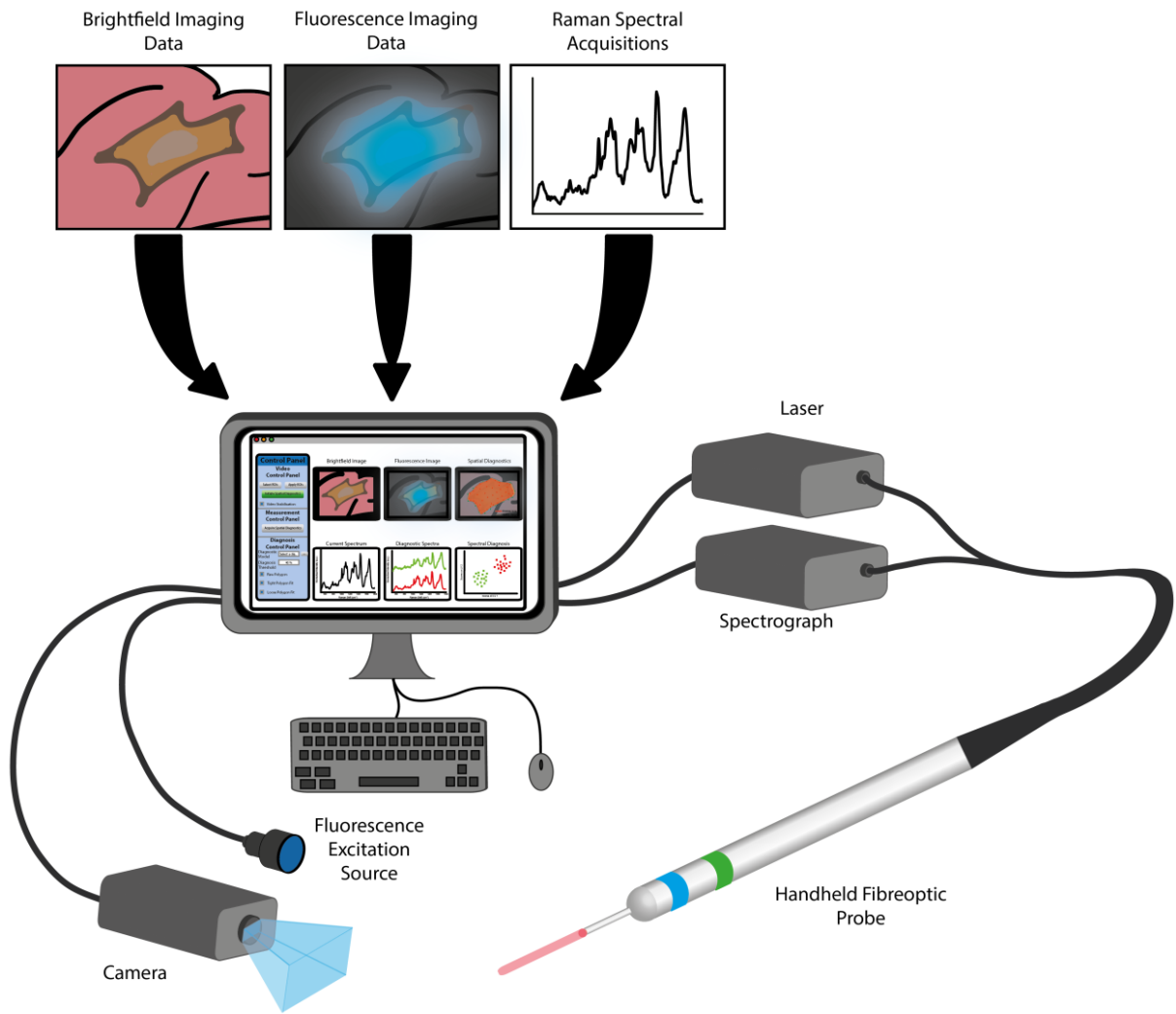

**Figure 1 | Fluorescence-Guided Raman Spectroscopic System.** Schematic of the fluorescence-guided Raman spectroscopic system comprising a camera for white light/fluorescence imaging, a handheld fibreoptic probe, a laser and spectrograph for Raman spectroscopy, an excitation light source, collection filter optics for fluorescence imaging, and a computer with integrated software for clinical control.

In the probe-tracking schema we developed, initial manual identification of the coloured, fiducial probe markers in the FOV enables HSV (Hue, Saturation, Value)-based image segmentation (Supplementary Figure 1), which is then combined with *a priori* knowledge of the probe and marker geometry for ratiometric calculation of the pose of both the spectroscopic probe and the probe tip (point of measurement acquisition) (Figure 2, I-II). The HSV colour space is used here as it has been shown to be superior to the RGB colour space for surgical tool tracking due to its decoupling of the chromaticity and luminance components[34,35]. Continuous acquisition (Figure 2, III) overlays the tracked location of the probe tip onto the clinical imaging video and records Raman spectral signal in real-time. Upon user identification of a region of interest, the system (Figure 2, IV) records both the location of the probe tip and a Raman spectral signal for diagnosis. This Raman spectral signal is then diagnosed in real-time through application of a previously developed diagnostic model (e.g. partial least squares-discriminant analysis (PLS-DA)) (Figure 2, V) and the diagnoses displayed at the probe tip coordinates are overlayed onto the clinical imaging video. Positive diagnoses are then connected to form a boundary that outlines the lesion margin. Importantly, the algorithm operates independently on each input video frame, accessing any previously stored diagnostic measurements to update margin delineation at each step, to ensure robust performance following temporary occlusion of the spectroscopic probe (Supplementary Figure 2).

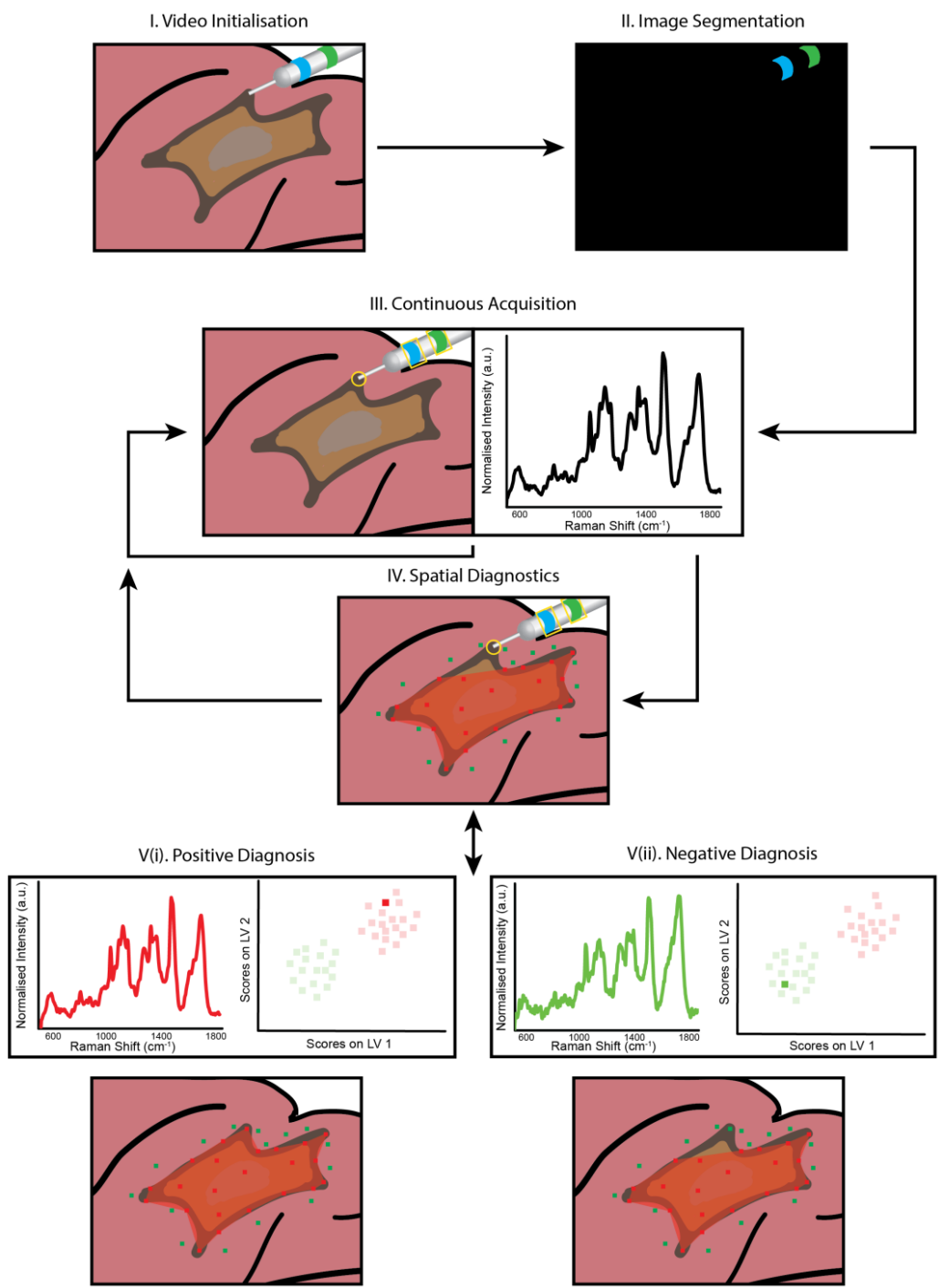

**Figure 2 | Fluorescence-Guided Raman Spectroscopic System Software Process Flow.** Schematic of fluorescence-guided Raman spectroscopic system software flow; **(I)** Video is initialised with diagnostic probe and interrogation area within FOV, **(II)** User interaction enables coloured-marker-based image segmentation which is combined with probe kinematic information for real-time probe tracking, **(III)** Once selected, the software begins near real-time data acquisition, continually tracking the diagnostic probe and recording spectral diagnostic information until the user starts a diagnostic acquisition, **(IV)** When the user starts a diagnostic acquisition, the coordinates of the probe are recorded and a detailed spectral signal acquired. This spectral signal is then diagnosed as either positive (**V, i**) or negative (**V, ii**) using a pre-developed spectroscopic diagnostic model. The diagnosis is then overlaid onto the imaging information at the coordinates where the measurement was acquired. Positive diagnosis coordinates are connected to form a boundary that delineates the tumour margin.

**White Light-Guided Spectroscopic Margin Delineation *Ex Vivo***

For *ex vivo* and *in vivo* applications, we implemented our spectroscopic margin delineation algorithm (Figure 2) within a clinician-facing graphical user interface (GUI) that provides functionality for system calibration, diagnostic spectroscopic model development, and margin delineation. The fluorescence-guided Raman spectroscopic system control software provides both a raw video input display and an augmented reality (AR) display of the surgical field of view, with spatially co-registered spectroscopic diagnostic information overlaid onto the raw video input.

Here, we examined the tissue discrimination capacity of our fluorescence-guided Raman spectroscopic system *ex vivo* using a combination of human squamous cell carcinoma (SCC) biopsy samples (n = 10 tissues) and human normal (muscle/non-cancerous and fatty) tissue samples obtained from abdominoplasty procedures (n = 4 tissues). Raman spectra were acquired using our fluorescence-guided Raman spectroscopic system for cancerous tissue (n = 201 spectra), normal (muscle/non-cancerous) tissue (n = 64 spectra), and normal (fatty) tissue (n = 89 spectra) with a 1 second acquisition time. Together, these spectra were used to develop a PLS-DA model discriminating between cancerous, normal (muscle/non-cancerous), and normal (fatty) tissue with cross-validated accuracies of 92.0%, 90.6%, and 95.7%, respectively (Supplementary Figure 3). Using this model, we were able to effectively delineate a region of cancerous tissue from surrounding normal (muscle/non-cancerous) and normal (fatty) tissue for a human squamous cell carcinoma biopsy sample, confirmed histologically using adjacent haematoxylin and eosin (H&E) stained sections (Figure 3, Supplementary Video 1). The system displays real-time spectroscopic information and automated PLS-DA diagnosis for each acquisition, providing an AR display of the specimen where the locations of negative (non-cancerous) acquisitions are displayed as green squares (with numbers indicating order of acquisition) and the locations of positive (cancerous) acquisitions are displayed as red squares (Figure 3a,d). The locations of positive diagnoses are then used to automatically delineate a predicted cancerous region, updated in real-time as subsequent diagnostic acquisitions are performed.

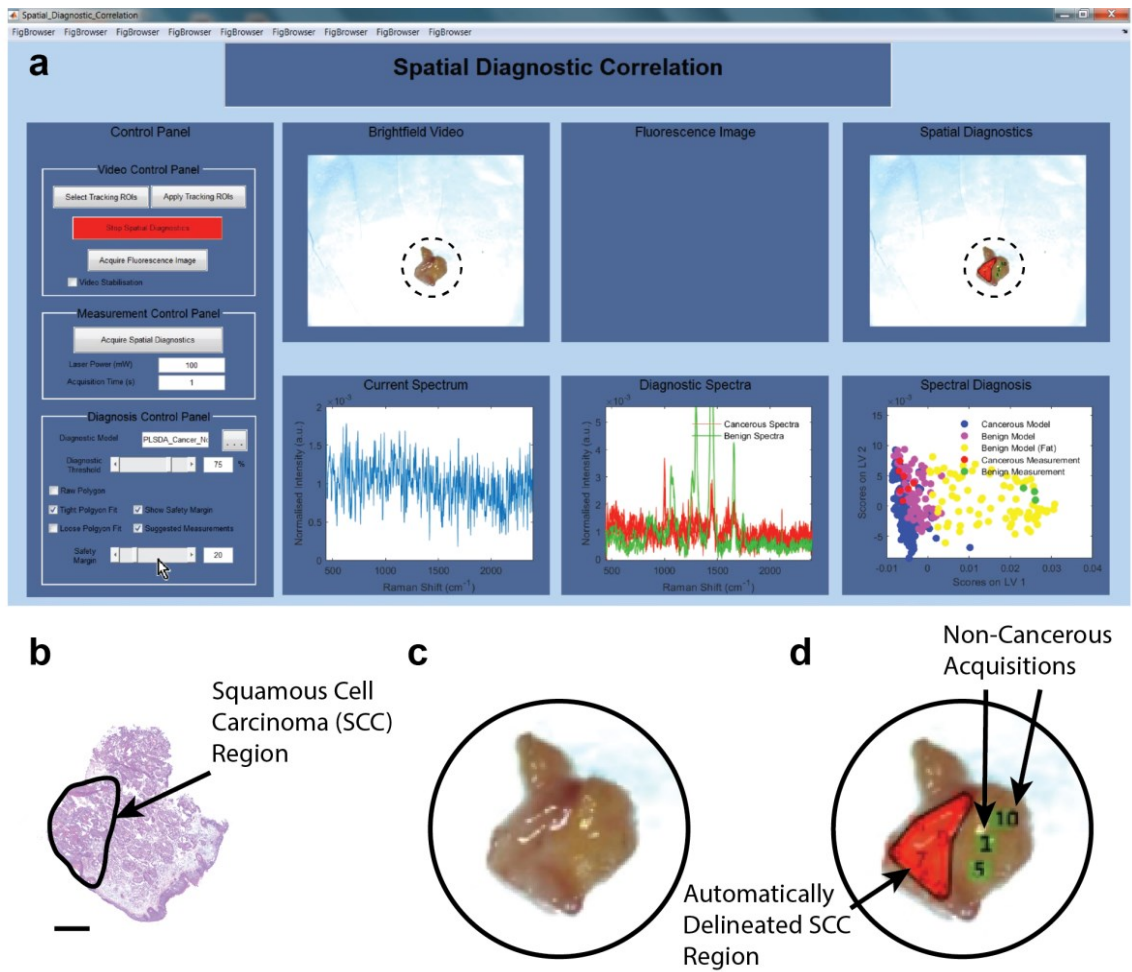

**Figure 3 | White Light-Guided Spectroscopic Margin Delineation *Ex Vivo*.** (**a**) Screenshot of the fluorescence-guided Raman spectroscopic system GUI during spectroscopic margin delineation of a human SCC biopsy specimen. This GUI provides the clinical interface between the surgeon and the system, providing a series of diagnostic controls as well as diagnostic spectral, statistical, and spatial information. (**b**) Adjacent hematoxylin and eosin (H&E) stained section of SCC biopsy specimen (scale bar = 2 mm). (**c**) White light image of the bulk SCC biopsy specimen. (**d**) White light image of the bulk tumour biopsy specimen with overlaid Raman spectral measurements and tumour margin delineation (red) where green squares indicate locations negative (non-cancerous) measurements, red squares (obscured under tumour margin delineation) indicate locations positive (cancerous) measurements, and the numbers inside each square indicate order of acquisition.

To further examine the potential for clinical operation of our system, we investigated the probe tracking and margin delineation accuracies as well as its computational performance (Supplementary Figure 4). Probe tracking accuracy, assessed in a mock *ex vivo* setting using fresh chicken muscle tissue, demonstrated a mean probe tip tracking error of $1.07 \pm 0.50$ mm for a 180-frame video sequence of input size 640 x 480 pixels with a working distance of 20 cm, in line with existing research and commercial optical tracking systems, which have reported errors of between 0.5 - 4 mm[36–39] (Supplementary Figure 4a,c,d). Similarly, computational performance, measured during a mock spatial diagnostic procedure, yielded a processing time of $0.21 \pm 0.03$ seconds during initial probe tracking (0.1 second Raman spectral integration time) (Supplementary Figure 4b, i). Acquisition of the first diagnostic acquisition required $3.91 \pm 0.11$ seconds (1 second Raman spectral integration time) (Supplementary Figure 4b, ii), with subsequent diagnostic acquisitions requiring just $1.90 \pm 0.13$ seconds (1 second Raman spectral integration time) (Supplementary Figure 4b, iii) due to MATLAB's just-in-time compilation[40]. Subsequent probe tracking processing time stabilised at $0.49 \pm 0.09$ seconds (0.1 second Raman spectral integration time) (Supplementary Figure 4b, iv).

While probe tracking here is performed in near real time (~2-5 frames per second (fps)) using a laptop, implementation on a more powerful computer together with code parallelisation would likely enable real time performance. Indeed, real time (> 30 fps) surgical tool tracking has been demonstrated using similar visual coloured marker-based tracking approaches (~55 fps)[41] and more recently with fully convolutional neural network deep learning strategies (~30 fps)[42]. The most significant impact on computational performance in our system is the integration time required for the acquisition of Raman spectral data (0.1 seconds for continuous acquisitions during probe movement and 1 second for diagnostic acquisitions). In each case, the integration times applied could likely be significantly reduced (by a factor of up to 10x) as *in vivo* Raman spectroscopic diagnostics has been achieved using integration times as low as 0.1 seconds[19,43,44].

To support clinical operation, we implemented several features designed to enable users to tailor spatial spectroscopic diagnostics to a particular patient or lesion (Supplementary Figure 5, Supplementary Video 2). These features, comprising adjustable safety margins, automatic suggested measurement locations, adjustable diagnostic thresholds, and video stabilisation, were demonstrated using an *ex vivo* fresh chicken tissue specimen to enable visual confirmation of software performance. E*x vivo* chicken tissue delineation was performed using a PLS-DA model to delineate fatty tissue from muscle tissue (100% cross-validated accuracy) (Supplementary Figure 6). Each of these features is intended to maximise robustness of our system to different clinical settings and maintain clinician diagnostic control.

To quantify the accuracy of our fluorescence-guided Raman spectroscopic system we delineated fatty chicken tissue from surrounding muscle tissue, enabling comparison with ground truth visual margin delineation (Supplementary Figure 7). Using our previously developed PLS-DA model (Supplementary Figure 5), we evaluated the margin delineation accuracy of our fluorescence-guided Raman spectroscopic system for three *ex vivo* chicken tissue specimens. Delineated areas closely resembled the ground truth areas in each case, as determined via manual image segmentation (Supplementary Figure 7). Balancing true positive delineated area with false negative and false positive delineated areas, the best results were obtained using a safety margin of 10 pixels for this particular setup, with a mean delineated true positive area of 86.5%, false negative of 13.5%, and false positive of 21.7% (Supplementary Figure 8).

**White Light-Guided Spectroscopic Margin Delineation *In Vivo***

We next investigated the feasibility of *in vivo* spectroscopic margin delineation in a *nu/nu* mouse SW122 colorectal cancer xenograft tumour model (Figure 4). *In vivo* Raman spectra (n = 320) from SW122 xenograft tumours and control flanks of two mice were collected across multiple timepoints and used to develop a PLS-DA model with a cross-validated accuracy of 96.5% (Figure 4d, Supplementary Figure 9). Applying our fluorescence-guided Raman spectroscopic system together with this PLS-DA model enabled real-time spectroscopic margin delineation of the SW122 tumour tissue from the surrounding healthy tissue. Crucially this process, in which each spectral acquisition takes ~2 seconds, can be performed in as little as 1 to 2 minutes depending on the complexity of tumour geometry. While margin delineation accuracy in this case is much harder to calculate than for an *ex vivo* model due to the difficulties in obtaining ground truth data, this result points to the clinical potential of our fluorescence-guided Raman spectroscopic system for aiding tumour resection procedures in intraoperative settings. However, while promising, it should be noted that for such clinical applications, margin delineation accuracy is dependent on both the probe tracking accuracy and the accuracy of the diagnostic model itself, e.g. the PLS-DA model applied here.

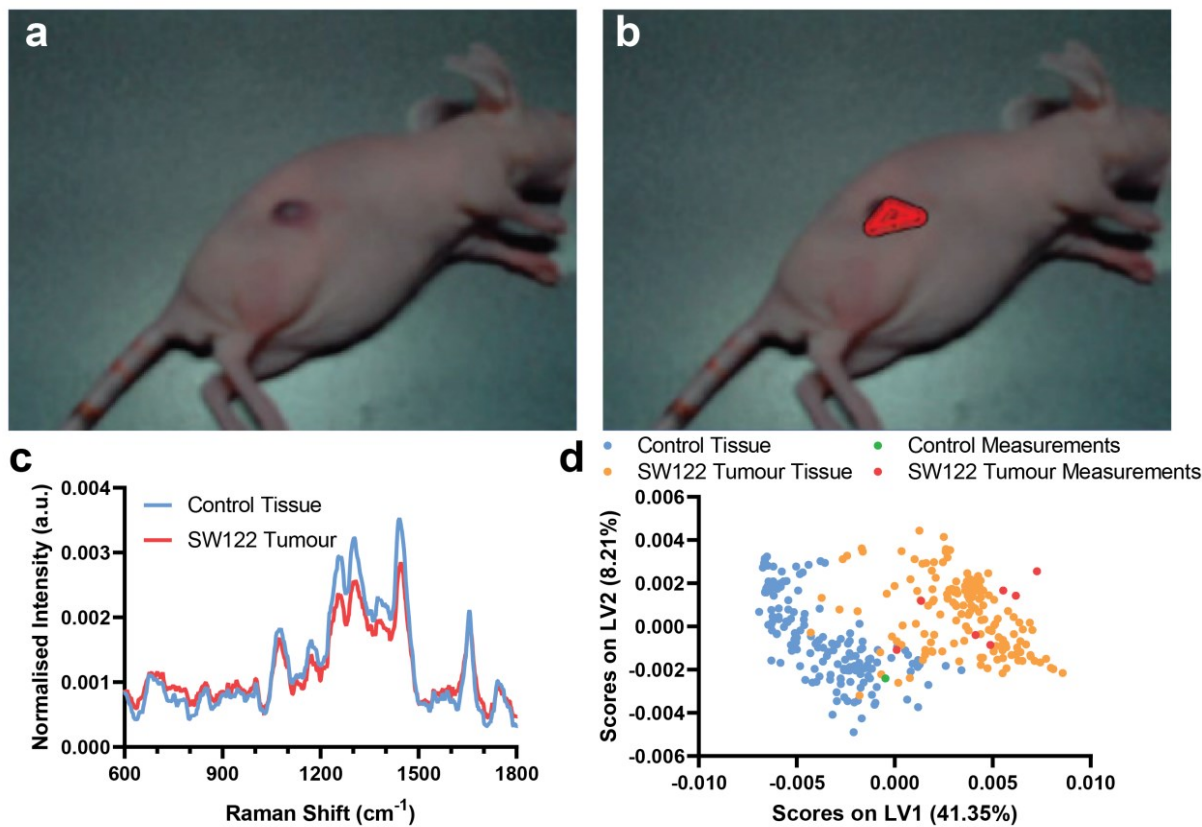

**Figure 4 | White Light-Guided Spectroscopic Margin Delineation *In Vivo*.** (**a-b**) Screenshots from the fluorescence-guided Raman spectroscopic system GUI during spatial spectroscopic diagnosis of (**a**) an SW122 colorectal xenograft tumour in a *nu/nu* mouse and (**b**) with AR Raman margin delineation overlay. (**c**) Mean Raman spectra of control tissue and SW122 tumours (N = 2, n = 80) used for PLS-DA. (**d**) PLS-DA latent variable 1 and 2 (LV1 and LV2) scores for control tissue and SW122 tumours.

**Fluorescence-Guided Spectroscopic Margin Delineation *Ex Vivo***

While the *in vivo* and *ex vivo* margin delineation results achieved with our system under white light guidance enabled the precise delineation of tumour margins, they do not address the first essential component for successful tumour resection – the initial detection and identification of suspicious lesions. Indeed, the macroscopic similarity between cancerous and healthy tissues is a key driver behind high post-surgical positive margin rates for many cancers. Therefore, to extend our system for comprehensive tumour resection assistance, we implemented fluorescence guidance for our Raman spectroscopic margin delineation.

After examining a range of fluorescent compounds currently employed for clinical and pre-clinical fluorescence-guidance applications, we selected the photosensitiser compound protoporphyrin IX (PPIX). As we have previously demonstrated[45], PPIX displays a low fluorescence background at clinically relevant concentrations under 785 nm Raman excitation (as typically applied in clinical Raman diagnostics) and is currently approved in the US and Europe for fluorescence-guided resection of high-grade gliomas[12,46–48].

In order to quantifiably demonstrate the benefit of fluorescence-guided Raman spectroscopic margin delineation, we developed a series of tissue-mimicking fluorescent optical phantoms containing cellular and serum components as well as varying PPIX concentrations (Supplementary Figure 10). Raman spectroscopic characterisation demonstrated small differences in background fluorescent intensity between the 0 μM, 2 μM, and 4 μM PPIX optical tissue phantoms with no obvious loss of Raman spectroscopic information, while the 20 μM optical tissue phantom demonstrated considerable fluorescence background and was readily distinguished from the remaining optical tissue phantoms via principal component analysis (PCA). Importantly, work quantifying the PPIX levels present in grade IV gliomas following the application of the PPIX precursor, 5-ALA, for fluorescence-guided surgery has indicated a mean concentration of 5.8 μM[49] and we have previously demonstrated the possibility of performing Raman spectroscopic diagnosis *in vivo* on PPIX-containing tissues[45].

We next developed a PLS-DA model to discriminate chicken muscle tissue from a non-fluorescent (0 μM PPIX) optical tissue phantom (100% cross-validated accuracy). We then applied this PLS-DA model for fluorescence-guided Raman margin delineation of both fluorescent and non-fluorescent optical tissue phantoms (Supplementary Figure 11). We were thus able to both detect fluorescent optical tissue phantoms within chicken tissue via fluorescence imaging and perform margin delineation via Raman spectroscopy (Figure 5). Although Raman spectral discrimination of the optical tissue phantoms and chicken tissue is trivial due to the significant spectral differences between the two materials, this experimental setup enabled a quantitative comparison of the margin delineation accuracy of our fluorescence-guided Raman spectroscopic system and fluorescence imaging alone for a series of 0 μM, 2 μM, and 4 μM PPIX optical tissue phantoms (Supplementary Figure 12). While Raman spectroscopic margin delineation accuracy remained constant across the three PPIX concentrations, fluorescence imaging margin delineation was much more dependent on the PPIX concentration, with fluorescence imaging unable to delineate the non-fluorescent optical tissue phantom (0 μM PPIX) (Supplementary Figure 13). When excluding the 0 μM PPIX optical tissue phantom data results, our fluorescence-guided Raman spectroscopic system performs as well as fluorescence imaging alone, with no statistically significant difference in the true positive, false negative, or false positive areas delineated by the two techniques (Student's t-test, n = 6) (Supplementary Figure 13c). However, our fluorescence-guided Raman spectroscopic system significantly outperformed fluorescence imaging alone for the 0 μM PPIX optical tissue phantom, with important implications for clinical applications to non-fluorescing tumour regions.

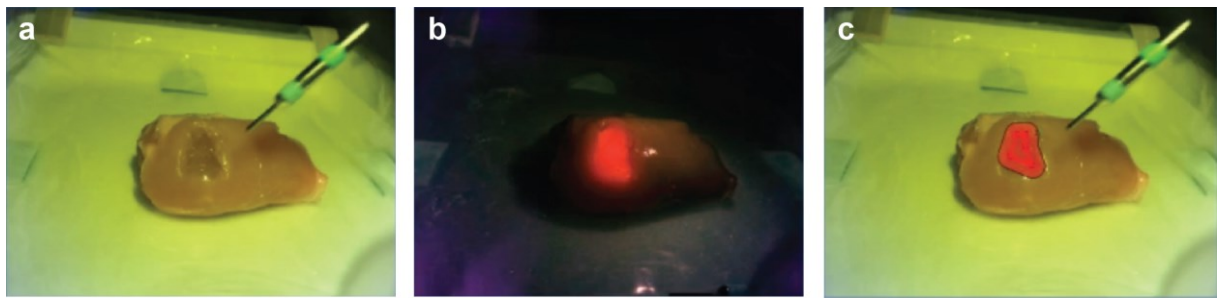

**Figure 5 | Fluorescence-Guided Spectroscopic Margin Delineation *Ex Vivo*. (a-c)** Screenshots of the fluorescence-guided Raman spectroscopic system during diagnosis and margin delineation of a 4 μM PPIX optical tissue phantom inserted into *ex vivo* chicken muscle tissue, (**a**) white light image, (**b**) fluorescence image, (**c**) AR overlay of Raman margin delineation onto white light image.

Next, to better assess the potential of our fluorescence-guided Raman spectroscopic system for tumour margin delineation, we created a composite optical tissue phantom with regions of varied fluorescence intensity (PPIX concentration) (Figure 6). Studies of FGS for high-grade gliomas have documented varying fluorescence intensity across tumours, particularly in necrotic cores, regions of occluded tissue, and, most importantly, at the tumour margins[12,50]. Here, margin delineation by our fluorescence-guided Raman spectroscopic system vastly outperformed fluorescence imaging alone (87% true positive diagnostic area with safety margin of 10 pixels vs 26 % for fluorescence imaging), with our system detecting the optical tissue phantom independent of PPIX concentration. In contrast, fluorescence imaging was only able to detect the 4 μM PPIX region, due to the occlusion of the 2 μM PPIX region under a thin layer (~ 1 mm) of 0 μM PPIX optical tissue phantom.

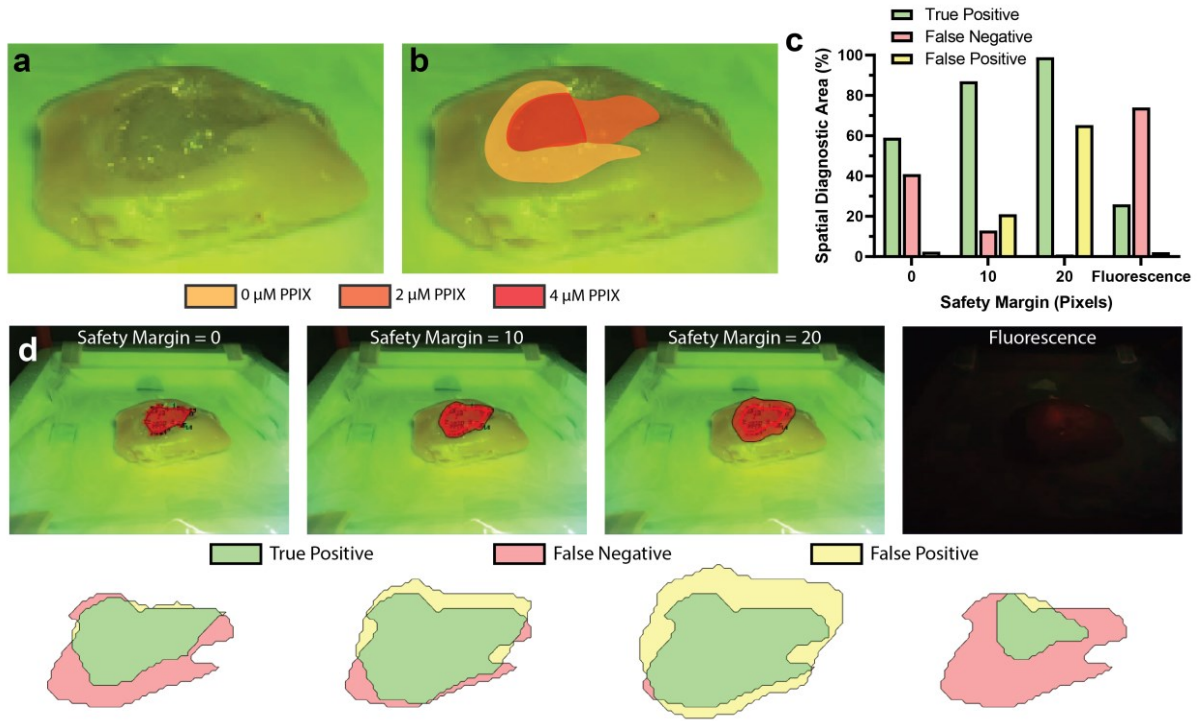

**Figure 6 | Tumour Mimicking Fluorescent Optical Tissue Phantom Margin Delineation Accuracy Evaluation.** (**a**) Photograph of the tumour mimicking fluorescent optical tissue phantom with regions of varying PPIX concentration. (**b**) Photograph of the tumour mimicking fluorescent optical tissue phantom with overlaid map of PPIX concentration regions. (**c**) Margin delineation accuracy for the tumour mimicking fluorescent optical tissue phantom via fluorescence-guided Raman spectroscopic margin delineation with safety margins of 0, 10, and 20 pixels, and via fluorescence imaging. (**d**) Margin delineation of tumour mimicking optical tissue phantom using fluorescence-guided Raman spectroscopic margin delineation (with safety margins of 0, 10, and 20 pixels) and fluorescence imaging. (**e**) Corresponding true positive, false negative, and false positive areas for fluorescence-guided Raman spectroscopic margin delineation (with safety margins of 0, 10, and 20 pixels) and fluorescence imaging.

This result thus demonstrates the complementarity of fluorescence imaging with Raman spectroscopic diagnostics, whereby fluorescence imaging enables rapid identification of fluorescence positive regions followed by spatial spectroscopic diagnosis for accurate margin delineation independent of variation in fluorescence intensity. As such, by developing diagnostic models that combine both fluorescence imaging information and Raman spectroscopic information the two modalities could yield more accurate margin delineation in combination than either modality would achieve in isolation. However, an important caveat to note is that margin delineation accuracy via our fluorescence-guided Raman spectroscopic system is dependent on both the probe tracking accuracy and the accuracy of the underlying diagnostic model for spectroscopic discrimination between tumour and healthy tissue. Given the significant spectral differences between the optical tissue phantoms and the chicken tissue used for these experiments, the results presented here reflect an idealised case where the underlying diagnostic model has a 100% discrimination accuracy.

## Discussion

Successful surgical treatment of tumours is highly dependent on the extent of tumour resection[51,52]. Despite this, post-surgical positive margin rates remain as high as 15-60% for a host of cancers due to both a lack of visual cues for the discrimination of healthy and cancerous tissues as well as the need to balance maximal tumour resection with healthy tissue preservation[53]. Though many technologies, such as intraoperative fluorescence imaging and spectroscopic diagnostics, have been developed towards improving tumour resection accuracy, no single system has adequately enabled both rapid tumour identification and accurate margin delineation[54]. The fluorescence-guided Raman spectroscopic system we report here is thus designed to enable both, without the introduction of complex, bulky, or expensive additional hardware into the operating theatre.

High information content optical spectroscopies such as Raman spectroscopies have long held promise for tumour margin delineation. However, the long acquisition times required have thus far restricted spectroscopy systems to point-based applications such that they cannot provide clinicians with a visual demarcation of a tumour boundary[4,31,55]. In contrast, fluorescence imaging offers visuals of a tumour's extent, though this has typically been limited to high-grade tumours that generate sufficient contrast[56]. Importantly, the performance of fluorescence imaging at tumour margins is typically low, impacting the ability to achieve accurate margin delineation[47].

Our fluorescence-guided Raman spectroscopic system is designed to bridge the gap between the high information content of spectroscopy and the visual-spatial information provided by white light/fluorescence imaging. Through tracking the pose of a spectroscopic probe during spectral acquisitions, our system enables an AR display of the surgical FOV, overlaying spatially co-registered spectroscopic diagnoses onto white light/fluorescence video image input for margin delineation. This combination of spatial and spectroscopic data thereby represents the first demonstration of rapid and accurate spectroscopic tumour margin delineation for both *ex vivo* and *in vivo* settings.

While our approach to tumour margin delineation has potential, several limitations will need to be overcome in order to enable successful clinical application. Firstly, performance should be enhanced to enable real-time operation. While much of this enhancement could be achieved through improved code parallelisation and computing hardware, this is also likely to require the use of a more sensitive Raman spectrometer in order to reduce signal acquisition times. Secondly, while the system currently provides a 2D AR overlay of spectroscopic and visual information, robust clinical application will likely necessitate extension to 3D through the use of multiple cameras for stereovision[57,58]. This would enable improved depth perception for both probe tracking and margin delineation that could better represent complex tumours contours *in vivo*. Finally, for fluorescence-guided approaches, care will need to be taken to develop diagnostic models that include data from tumours with varying fluorescence intensities in order to maximise performance[59]. While we have demonstrated fluorescence-guided Raman spectroscopy here using only optical tissue phantoms, we have previously reported on the feasibility Raman spectroscopic diagnosis of PPIX-containing tissues *in vivo* and *ex vivo*[45]. Combined, these developments would likely prepare our prototype system for more comprehensive preclinical and clinical testing to evaluate its benefits for tumour resection surgeries.

In conclusion, the combination of Raman spectroscopic diagnosis with fluorescence imaging and computer vision probe tracking thus opens exciting opportunities for clinical tumour resection guidance. Given the improved cancer detection sensitivity spectroscopic methods have consistently shown relative to white light/fluorescence imaging modalities, our fluorescence-guided Raman spectroscopic system will likely enable more accurate tumour margin delineation and could lead to improved rates of complete resection and thus better patient outcomes.

## Materials and Methods

### Fluorescence-Guided Raman Spectroscopic System

The fluorescence-guided Raman spectroscopic system consisted of seven key components; a Raman spectroscopic probe with a 2.1 mm diameter (EmVision), a spectrograph (QEPro, OceanOptics), a Raman spectroscopy laser source (785 nm, 600 mW, B&W Tek), a 405 nm LED light source (600 mW, Thorlabs), an 8 megapixel iSight camera (Apple) with a 500 nm longpass filter (Edmund Optics), and a computer (Lenovo Thinkpad T460, Intel Core i5-6200U CPU). System control software was developed in the MATLAB 2017B environment, using the MATLAB graphical user interface development environment (GUIDE), and the MATLAB image processing and computer vision toolboxes.

### Fluorescence-Guided Raman Spectroscopic System Characterisation

Probe tracking accuracy was calculated using a video of probe movement in an *ex vivo* margin delineation setting. Computed coordinate locations for each of the fiducial markers and the probe tip were compared to ground truth coordinates, generated as the mean (n = 3) user-identified coordinates for each video frame. Mean computation time (n = 3) was determined using the in-built MATLAB functions *tic* and *toc* for the core algorithm processing loop during a typical *ex vivo* margin delineation procedure.

### Ethics Statements

Human samples used in this research project were obtained from the Imperial College Healthcare Tissue Bank (ICHTB). ICHTB is supported by the National Institute for Health Research (NIHR) Biomedical Research Centre based at Imperial College Healthcare NHS Trust and Imperial College London. ICHTB is approved by Wales REC3 to release human material for research (17/WA/0161), and the samples for this project (R19022) were issued from the ICHTB Collection.

All animal studies were approved by the University College London Biological Services Ethical Review Committee and licensed under the UK Home Office regulations and the Guidance for the Operation of Animals (Scientific Procedures) Act 1986 (Home Office, London, United Kingdom) and United Kingdom Coordinating Committee on Cancer Research Guidelines for the Welfare and Use of Animals in Cancer Research[60].

### *Ex Vivo* Diagnostic Model Development

Using the fluorescence-guided Raman spectroscopic system, 25-30 Raman spectra were collected from raw *ex vivo* chicken muscle tissue, raw ex vivo chicken fat tissue, and the 0 µM, 2 µM, 4 µM, and 20 µM PPIX optical tissue phantoms. Fresh chicken tissue was obtained from a local butcher on the day of experiments. Spectral processing was performed in MATLAB. Spectra were cropped, background subtracted (Whittaker filter, $\lambda = 100,000$), normalised to the area under the curve, and filtered (Savitzky-Golay, $1^{st}$ order, frame width = 7) before different PLS-DA classification models were developed with Venetian blinds cross validation using PLS Toolbox (Eigenvector Research) within the MATLAB environment.

### *Ex Vivo* Spectroscopic Margin Delineation

*Ex vivo* fluorescence-guided Raman spectroscopic margin delineation was performed using the fluorescence-guided Raman spectroscopic system with a 785 nm laser at 100 mW power output and a 1 second integration time with a previously developed PLS-DA model applied prospectively.

### Histology

10 µm cryosections of human tissue biopsy samples were obtained following water embedding and freezing using a Bright OTF cryostat. Cryosections were subsequently formalin fixed and stained with haematoxylin and eosin (H&E) in triplicate for each sample. Histology sample preparation and staining was performed with the assistance of Lorraine Lawrence at Imperial College London's Research Histology Facility within the Facility for Imaging by Light Microscopy (FILM). Imaging of stained sections was performed using a Zeiss Axio Observer widefield inverted microscope with a 20x objective.

### *Ex Vivo* Margin Delineation Accuracy Characterisation

*Ex vivo* fluorescence-guided Raman spectroscopic margin delineation was performed on 3 raw chicken tissue specimens using the fluorescence-guided Raman spectroscopic system with a 785 nm laser at 100 mW power output and a 1 second integration time and a previously developed PLS-DA model applied prospectively. For each specimen, between 22-26 spectral acquisitions were obtained and used to delineate the chicken fat tissue. Algorithm-delineated areas of chicken fat tissue, with safety margin sizes of 0, 10, and 20 pixels, were compared to the ground truth area and the size of true positive, false negative, and false positive regions determined.

### *In Vivo* Spectroscopic Margin Delineation

Two female *nu/nu* mice (12 weeks old, 25-30g) were subcutaneously injected with $1 \times 10^6$ cells from a human colorectal carcinoma cell line, SW122, on their right flank. Using the fluorescence-guided Raman spectroscopic system, 80 Raman spectra were collected from control mouse tissue *in vivo* and 80 spectra from the SW122 xenograft tumours *in vivo* at 12 days after implantation. Spectra were cropped, background subtracted (Whittaker filter, $\lambda = 100,000$), normalised, and filtered (Savitzky-Golay, $1^{st}$ order, frame width = 7) before a PLS-DA classification model of the spectra was developed with a Venetian blinds cross validation using PLS Toolbox (Eigenvector Research) within the MATLAB environment. *In vivo* fluorescence-guided Raman spectroscopic margin delineation of the SW122 xenograft tumours was then performed using the fluorescence-guided Raman spectroscopic system with a 785 nm laser at 100 mW power output and a 1 second integration time using the previously developed PLS-DA model applied prospectively.

### Cell Culture and Preparation for Optical Tissue Phantoms

MDA-MB-231 cells (ATCC) for optical tissue phantoms were grown at 37 °C and 5% $CO_2$ in high glucose (4.5 g/L) DMEM GlutaMax (Life Technologies) supplemented with 10% (v/v) fetal bovine serum (FBS), 1x penicillin-streptomycin, 1x non-essential amino acids, and 20 mM pH 7.3 HEPES buffer solution. The cell line was authenticated using STR profiling. Prior to inclusion in optical tissue phantoms, cells were trypsinised, spun down at 300 x g for 5 minutes, washed, and then fixed in 4% (v/v) PFA in PBS for 20 minutes.

### Optical Tissue Phantom Construction

Optical tissue phantoms that mimic tissue absorption and scattering were prepared using an adaptation of a previously established protocol[61], by combining agarose (Sigma-Aldrich), water, PBS (ThermoFisher), foetal bovine serum (FBS) (ThermoFisher), homemade intralipid, and human haemoglobin. Homemade intralipid was made by forming a solution of 20% (v/v) sunflower oil and 1% DSPC (Avanti Polar Lipids) in water. Briefly, 200 mg of agarose was dissolved in a 10 mL 1:1 solution of $H_2O$:PBS and heated at 100 °C under constant stirring until a transparent solution was formed. Solution was then allowed to cool slowly to 50 °C. Once the solution reached 50 °C, 2.6 mL FBS, 200 μL of homemade intralipid solution, 200 μL of human haemoglobin, and 1 mL of 20 x $10^6$ fixed MDA-MB-231 cells/mL were added and stirred through the solution. For fluorescence-guided applications, varying amounts of 500 μM PPIX (Sigma-Aldrich) was also added to create 2 μM, 4 μM, and 20 μM PPIX fluorescent optical tissue phantoms. The solution was then poured into a mould (e.g. cavity cut out of chicken tissue), as required, and allowed to cool to room temperature to set before use.

## *Ex Vivo* Fluorescence-Guided Margin Delineation Accuracy Characterisation

*Ex vivo* fluorescence-guided Raman spectroscopic margin delineation was performed on raw chicken tissue specimens with 0, 2, or 4 μM PPIX optical tissue phantom inserts using the fluorescence-guided Raman spectroscopic system with a 785 nm laser at 100 mW and a 1 second integration time and a previously developed PLS-DA model applied prospectively. For each specimen, between 10-30 spectral acquisitions were obtained and used to delineate the fluorescent optical tissue phantoms. Algorithm-delineated areas of fluorescent optical tissue phantom, with safety margin sizes of 0, 10, and 20 pixels, were compared to the ground truth area and the size of true positive, false negative, and false positive regions determined.

## Acknowledgements

C.C.H. acknowledges funding from the NanoMed Marie Skłodowska-Curie ITN from the H2020 programme under grant number 676137. M.S.B. acknowledges support from H2020 through the Individual Marie Skłodowska-Curie Fellowship "IMAGINE" (701713). M.Z.T. and T.L.K acknowledge funding from the EPSRC for the award of an Early Career Fellowship EP/L006472/1. A.N. and M.M.S. acknowledge support from the GlaxoSmithKline Engineered Medicines Laboratory. D.J.S acknowledges support from a British Heart Foundation Intermediate Basic Science Research Fellowship (FS/15/33/31608), the MRC MR/R026416/1 and the Wellcome Trust. M.M.S. acknowledges a Wellcome Trust Senior Investigator Award (098411/Z/12/Z).

# Supplementary Information

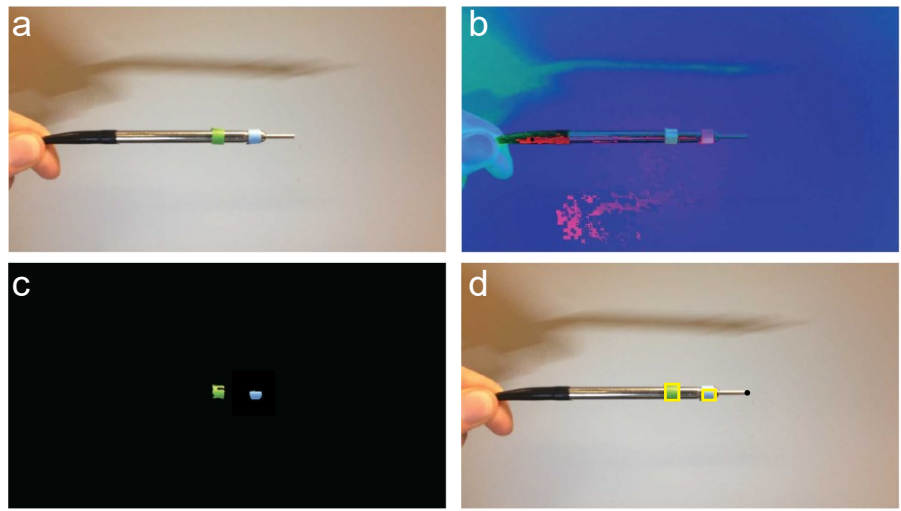

**Supplementary Figure 1 | HSV Image Segmentation for Coloured Marker-Based Spectroscopic Probe Tracking. (a-d)** Image processing for coloured marker-based tracking takes (**a**) an input image frame and converts it from an RGB image to (**b**) an HSV image before (**c**) HSV-based image segmentation for identification and isolation of the coloured fiducial markers, which (**d**) is then combined with a priori knowledge of the probe geometry to calculate the pose of the spectroscopic probe and the location of the probe tip (black circle).

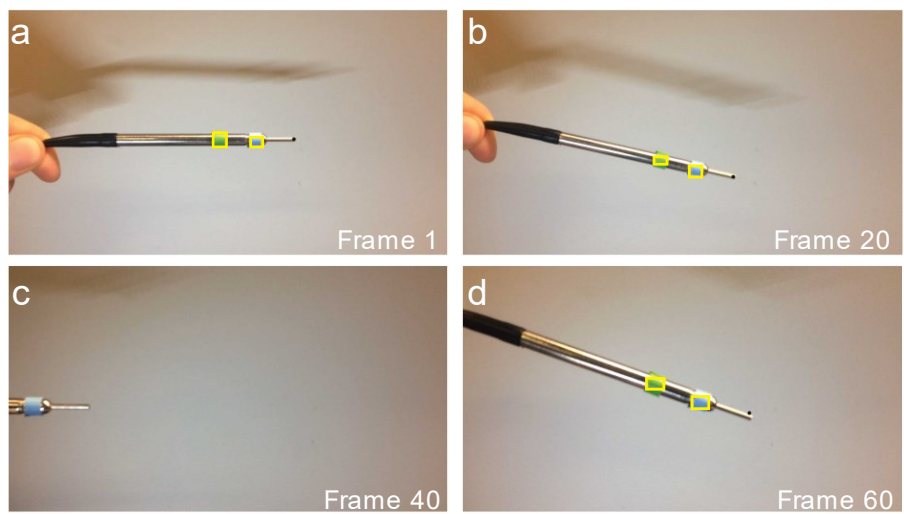

**Supplementary Figure 2 | Coloured Marker-Based Tracking of Spectroscopic Probe. (a-d)** Sequential coloured marker-based tracking video frames of the spectroscopic probe during a video sequence, with identified fiducial markers and probe tip, demonstrating robust tracking performance following partial occlusion of the spectroscopic probe.

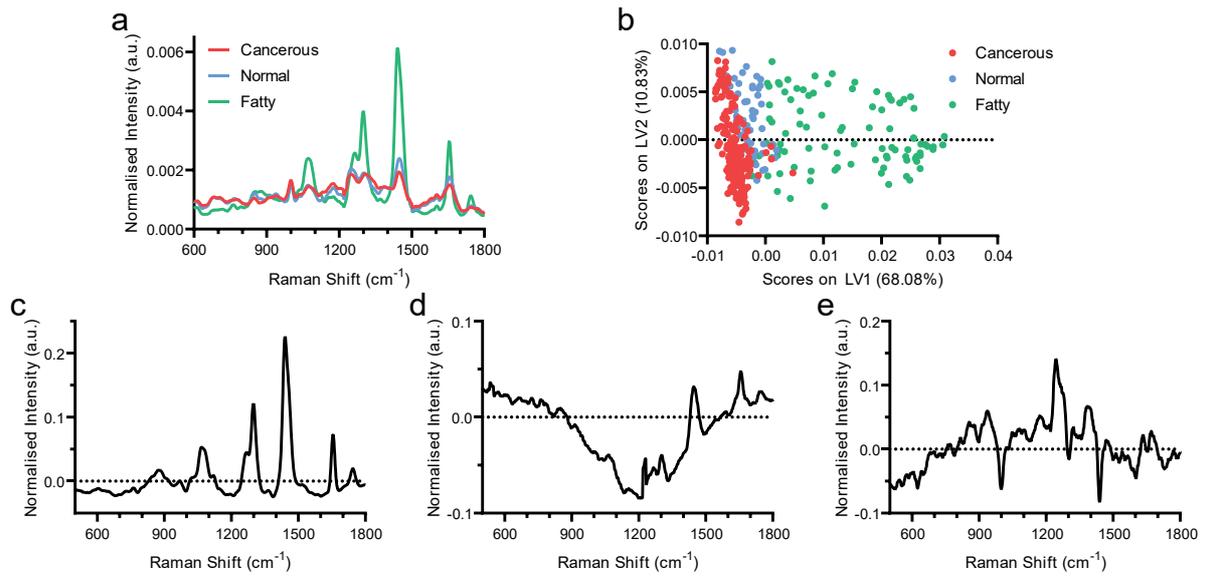

**Supplementary Figure 3 | Raman Spectroscopic PLS-DA Model of *Ex Vivo* Human Biopsy Specimens.** (**a**) Mean Raman spectra of cancerous, normal (non-fat), and fatty (normal) tissues (N = 3-4 tissues, n ≥ 20 spectra). (**b**) PLS-DA latent variable 1 and 2 (LV1 and LV2) scores for cancerous, normal (non-fat), and fatty (normal) tissues, where PLS-DA latent variables represent spectral features of descending importance that best enable separation of the different tissue classes. (**c**) PLS-DA latent variable 1. (**d**) PLS-DA latent variable 2. (**e**) PLS-DA latent variable 3.

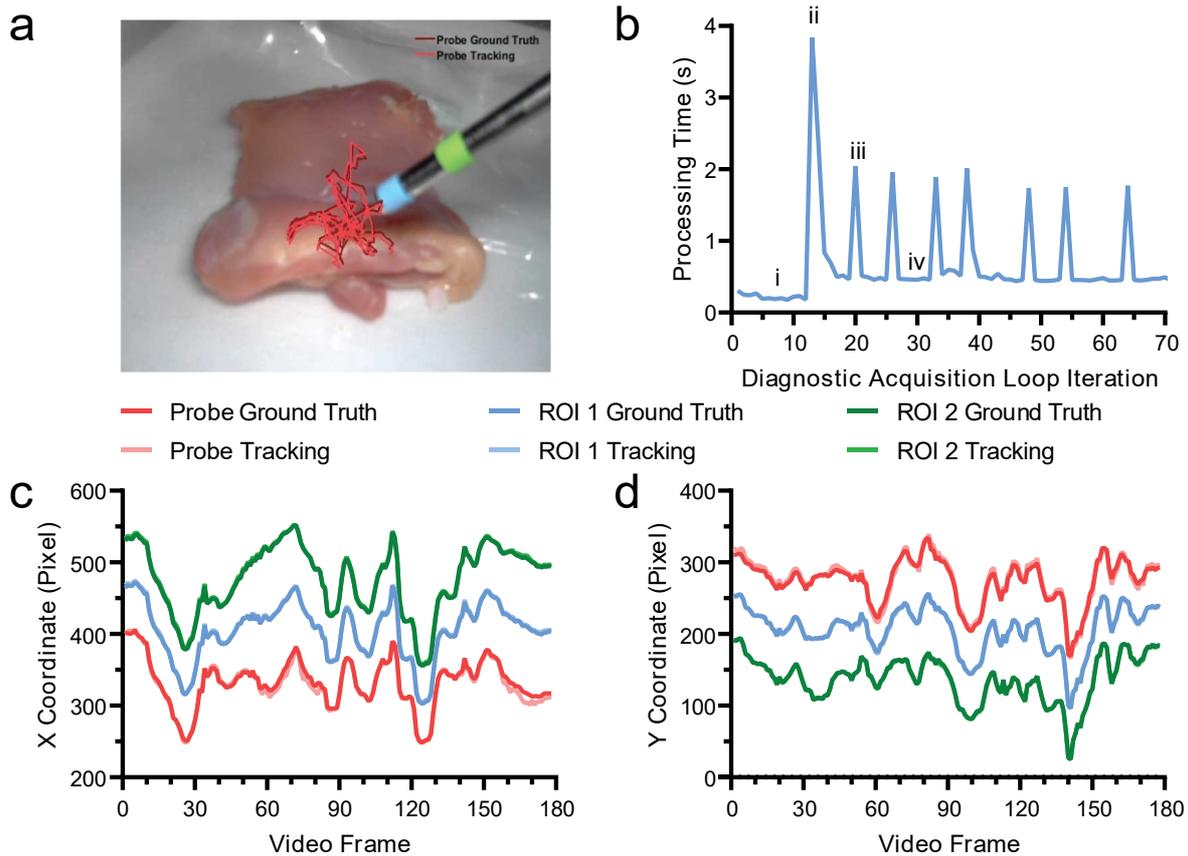

**Supplementary Figure 4 | Fluorescence-Guided Raman Spectroscopic System Characterisation.** (**a**) Video frame of mock Raman spectroscopic margin delineation video sequence of fresh chicken tissue used for characterisation of probe tracking error with overlaid ground truth and algorithm-determined probe-tip motion for entire video sequence. (**b**) Analysis of the processing time for the core spectroscopic margin delineation algorithm during (**i**) baseline loop iteration with 0.1 second integration time (prior to spectroscopic diagnostic acquisitions), (**ii**) initial spectroscopic diagnostic acquisition with 1 second integration time, (**iii**) subsequent spectroscopic diagnostic acquisitions with 1 second integration time, and (**iv**) baseline loop iterations with 0.1 second integration time (between spectroscopic diagnostic acquisitions) (n = 3). (**c-d**) Tracking error for probe tip and fiducial markers in (**c**) X and (**d**) Y for the mock spectroscopic margin delineation video sequence.

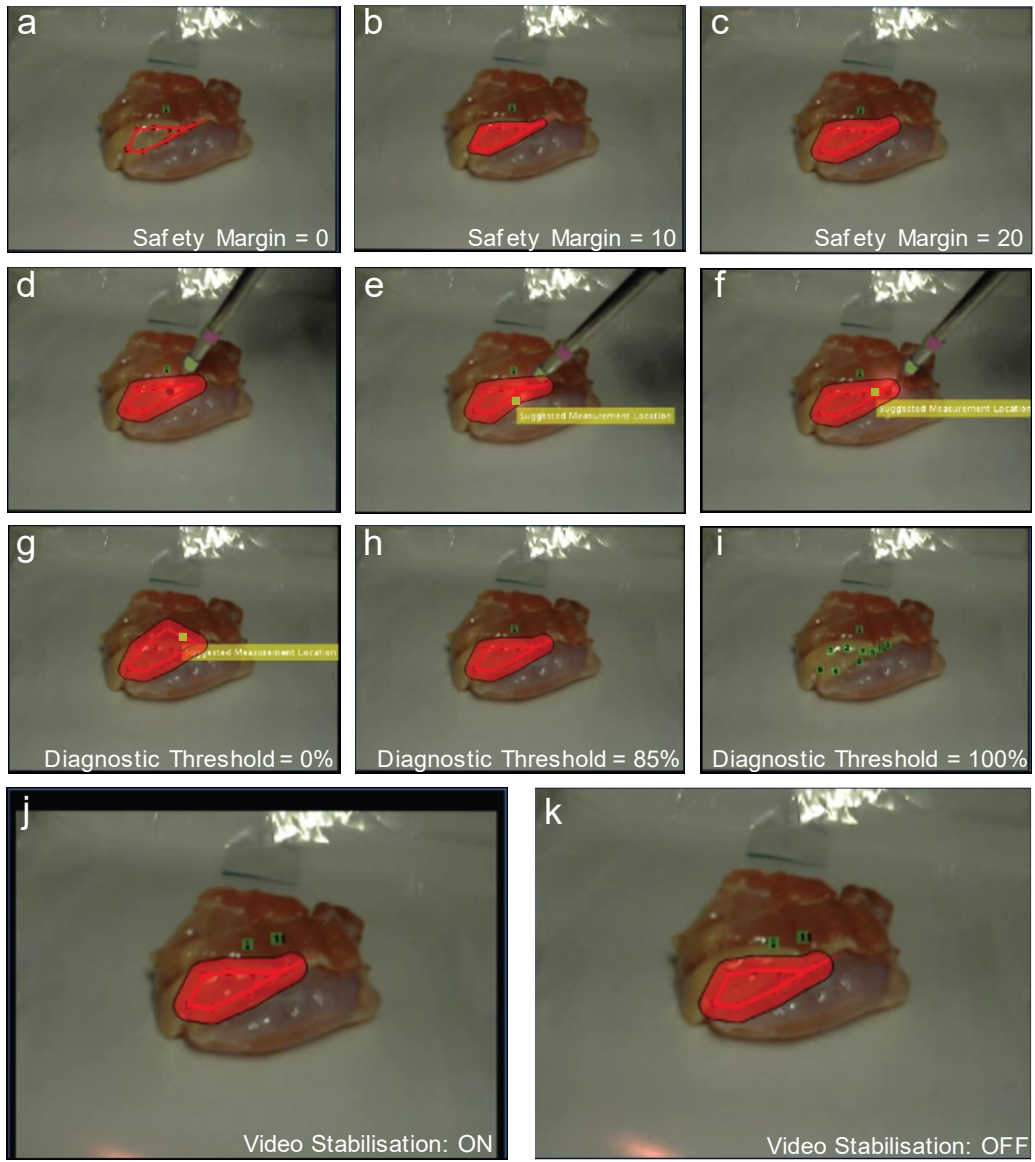

**Supplementary Figure 5 | Fluorescence-Guided Raman Spectroscopic System Features. a-c,** Video frames from the fluorescence-guided Raman spectroscopic system showing the delineated region of chicken fat tissue with safety margins value of (**a**) 0, (**b**) 10, and (**c**) 20 pixels. (**d-f**) Sequential video frames from the fluorescence-guided Raman spectroscopic system indicating iterative suggested measurement locations (yellow squares) dependent on the spatial distribution of existing positive diagnostic acquisitions. (**g-i**) Video frames from the fluorescence-guided Raman spectroscopic system showing the delineated region of chicken fat tissue with diagnostic thresholds of (**g**) 0%, (**h**) 85%, and (**i**) 100% where green squares indicate locations of negative (non-cancerous) acquisitions and red squares indicate positive (cancerous) acquisitions. (**j-k**) Video frames from the fluorescence-guided Raman spectroscopic system showing the resulting AR display of spatial spectroscopic diagnostic coordinates when video stabilisation is (**j**) enabled and (**k**) disabled.

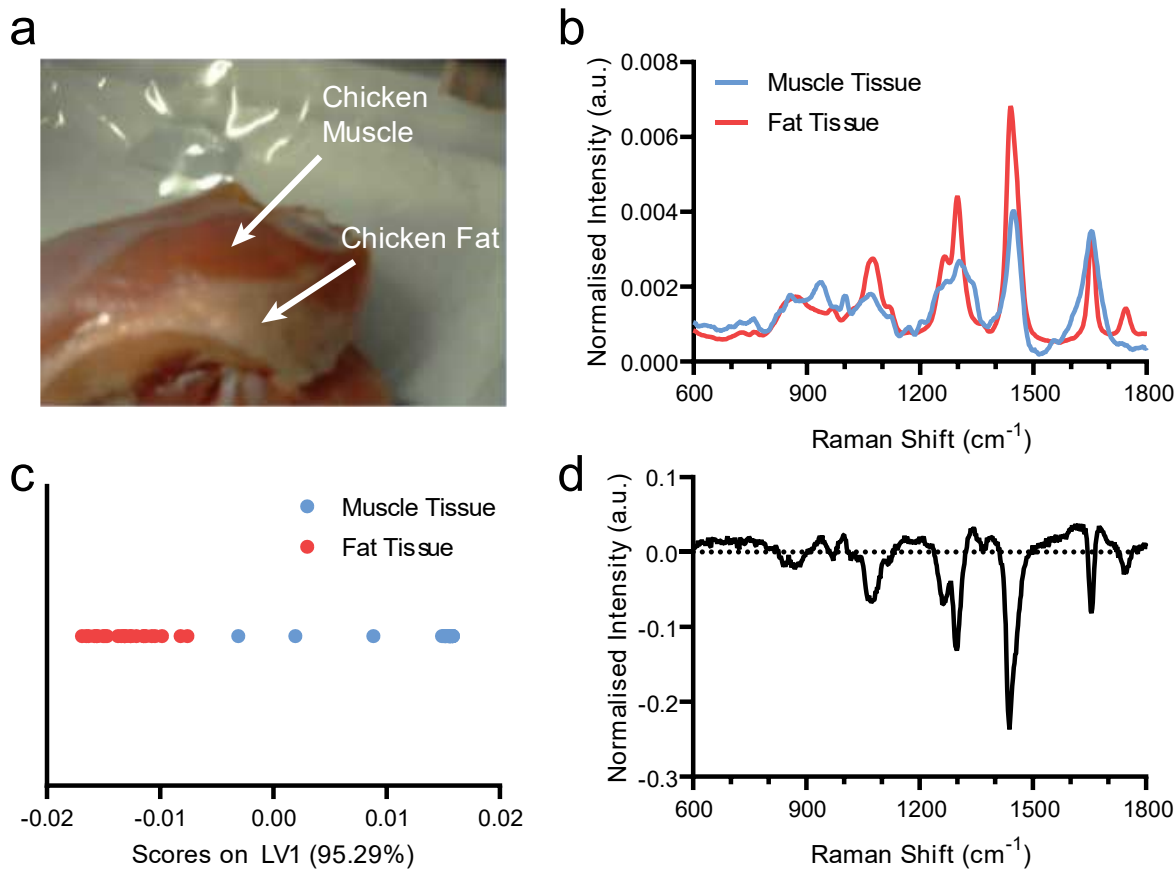

**Supplementary Figure 6 | Raman Spectroscopic PLS-DA Model of *Ex Vivo* Chicken Tissue.** (**a**) Video frame from the fluorescence-guided Raman spectroscopic system showing chicken muscle and chicken fat tissue. (**b**) Mean Raman spectra of chicken muscle tissue and chicken fat tissue (n = 25). (**c**) PLS-DA latent variable 1 (LV1) scores for chicken muscle tissue and chicken fat tissue Raman spectra. (**d**) PLS-DA latent variable 1.

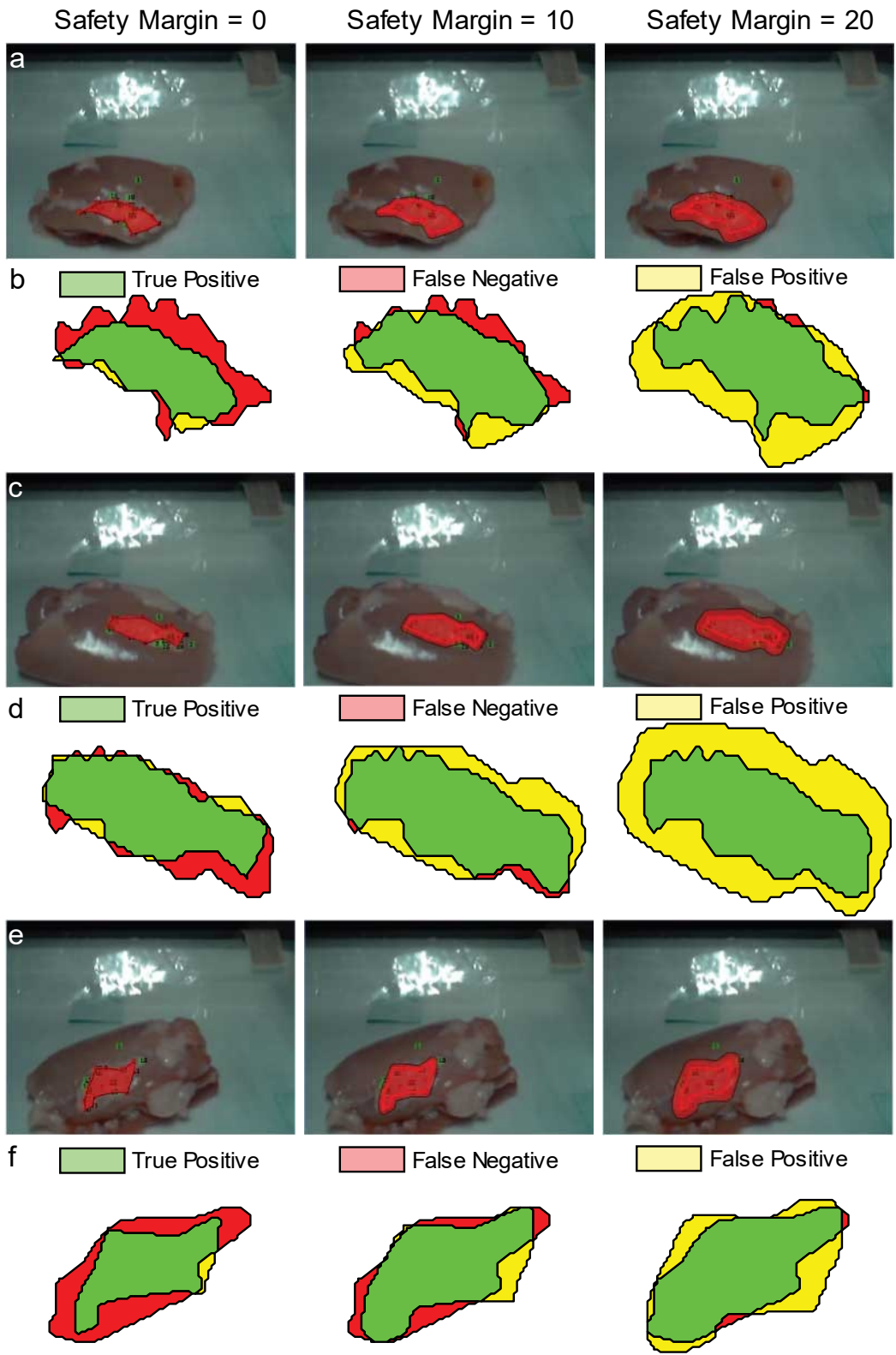

**Supplementary Figure 7 | White Light-Guided Spectroscopic Margin Delineation *Ex Vivo* Accuracy Evaluation. (a,c,e)** Video frames from our fluorescence-guided Raman spectroscopic system showing the delineated margin of chicken fat tissue following 22-26 diagnostic spectral acquisitions with safety margin sizes of 0, 10, and 20 pixels for (**a**) specimen 1, (**c**) specimen 2, and (**e**) specimen 3. (**b,d,f**) Corresponding margin delineation accuracies indicating true positive (green), false negative (red), and false positive (yellow) diagnostic regions for (**b**) specimen 1, (**d**) specimen 2, and (**f**) specimen 3.

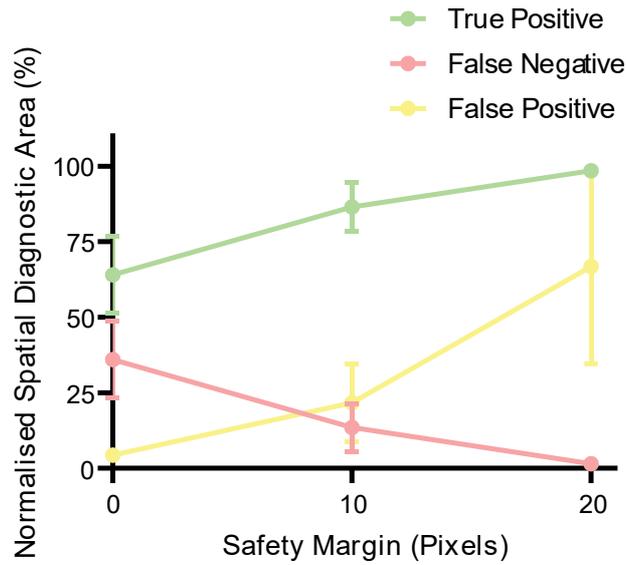

**Supplementary Figure 8 | White Light-Guided Spectroscopic Margin Delineation *Ex Vivo* Accuracy Quantification.** Mean normalised true positive, false negative, and false positive margin delineated areas for the *ex vivo* chicken tissue specimens.

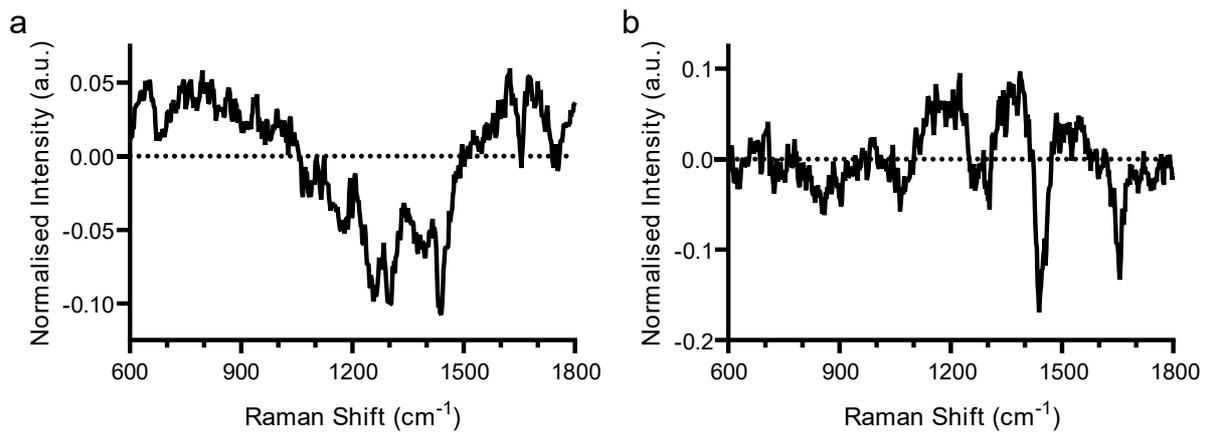

**Supplementary Figure 9 | Raman Spectroscopic PLS-DA Model Latent Variables for *In Vivo* SW122 Xenograft Mouse Tumour Model.** (**a**) PLS-DA latent variable 1. (**b**) PLS-DA latent variable 2.

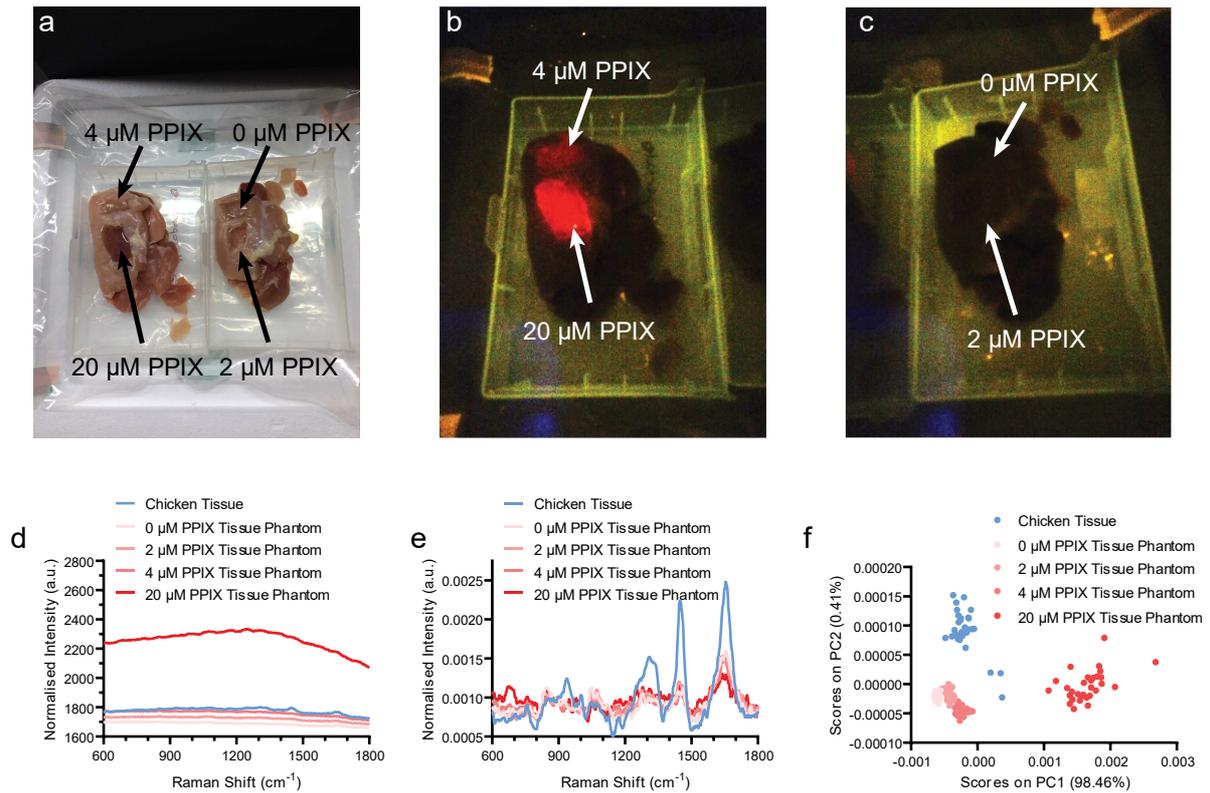

**Supplementary Figure 10 | Fluorescence Imaging and Raman Spectroscopy of PPIX Optical Tissue Phantoms in *Ex Vivo* Chicken Tissue.** (**a**) Photograph of 0 µM, 2 µM, 4 µM, and 20 µM PPIX optical tissue phantoms inserted into chicken muscle tissue. (**b-c**) Fluorescence imaging of 0 µM, 2 µM, 4 µM, and 20 µM PPIX optical tissue phantoms in *ex vivo* chicken tissue. (**d**) Raw and (**e**) processed Raman spectra of the PPIX optical tissue phantoms and *ex vivo* chicken muscle tissue (n = 30). (**f**) PCA of the PPIX optical tissue phantoms and *ex vivo* chicken tissue Raman spectra.

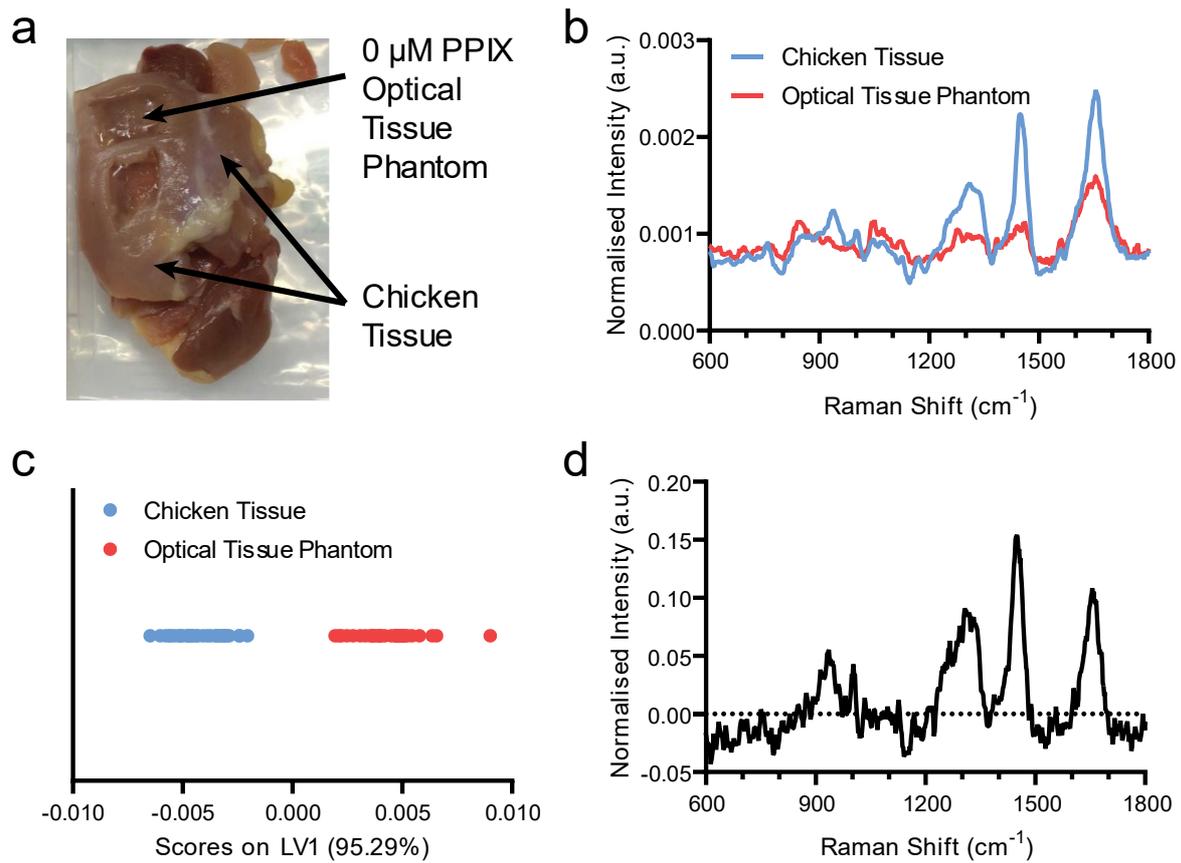

**Supplementary Figure 11 | Raman Spectroscopic PLS-DA Model of *Ex Vivo* Chicken Tissue and Optical Tissue Phantom.** (**a**) Photograph of the 0 μM PPIX optical tissue phantom inserted into chicken tissue used to generate the PLS-DA model. (**b**) Mean Raman spectra of chicken tissue and optical tissue phantom (n = 30). (**c**) PLS-DA latent variable 1 (LV1) scores for the chicken tissue and the optical tissue phantom Raman spectra. (**d**) PLS-DA latent variable 1.

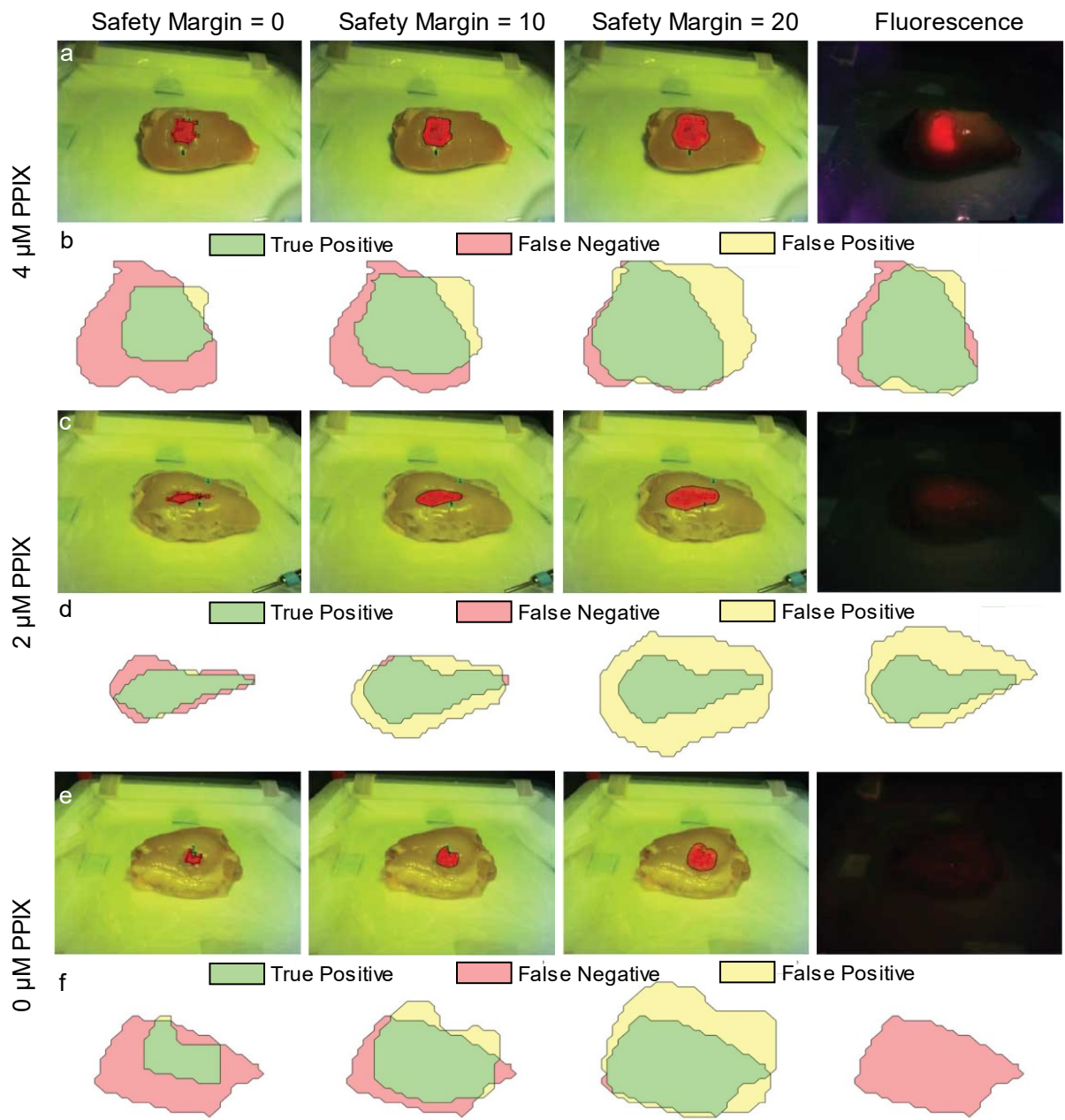

**Supplementary Figure 12 | Fluorescence-Guided Spectroscopic Margin Delineation *Ex Vivo* Accuracy Evaluation of PPIX Optical Tissue Phantoms.** (**a,c,e**) Exemplar margin delineation of (**a**) 4 µM PPIX, (**c**) 2 µM PPIX, and (**e**) 0 µM PPIX optical tissue phantoms using fluorescence-guided Raman spectroscopic margin delineation (with safety margins of 0, 10, and 20 pixels) and fluorescence imaging. (**b,d,f**) Corresponding true positive, false negative, and false positive areas for fluorescence-guided Raman spectroscopic margin delineation (with safety margins of 0, 10, and 20 pixels) and fluorescence imaging.

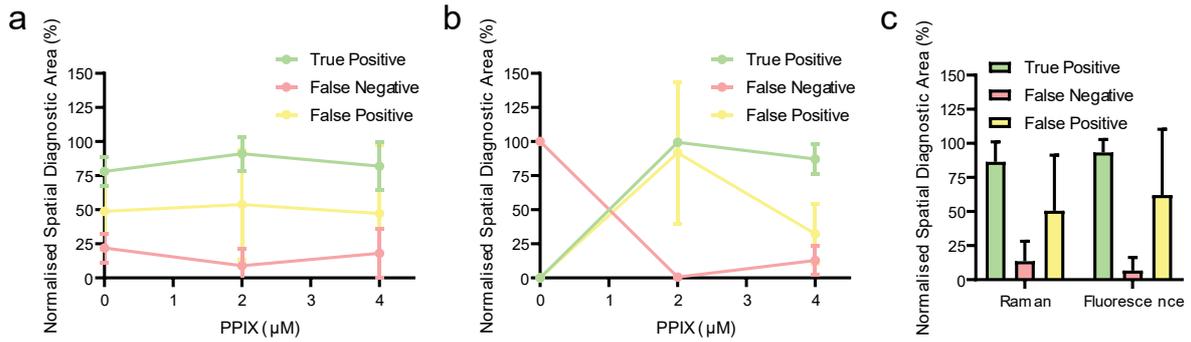

**Supplementary Figure 13 | Mean Fluorescence-Guided Spectroscopic Margin Delineation *Ex Vivo* Accuracy. (a-b)** Mean phantom margin delineation accuracy for the 0 μM, 2 μM, and 4 μM PPIX optical tissue phantoms (n = 3) using **(a)** fluorescence-guided Raman spectroscopic margin delineation and **(b)** fluorescence imaging. **(c)** Comparison of the mean margin delineation accuracy of fluorescence-guided Raman spectroscopic margin delineation and fluorescence imaging across the fluorescent optical tissue phantoms (2 μM and 4 μM PPIX).

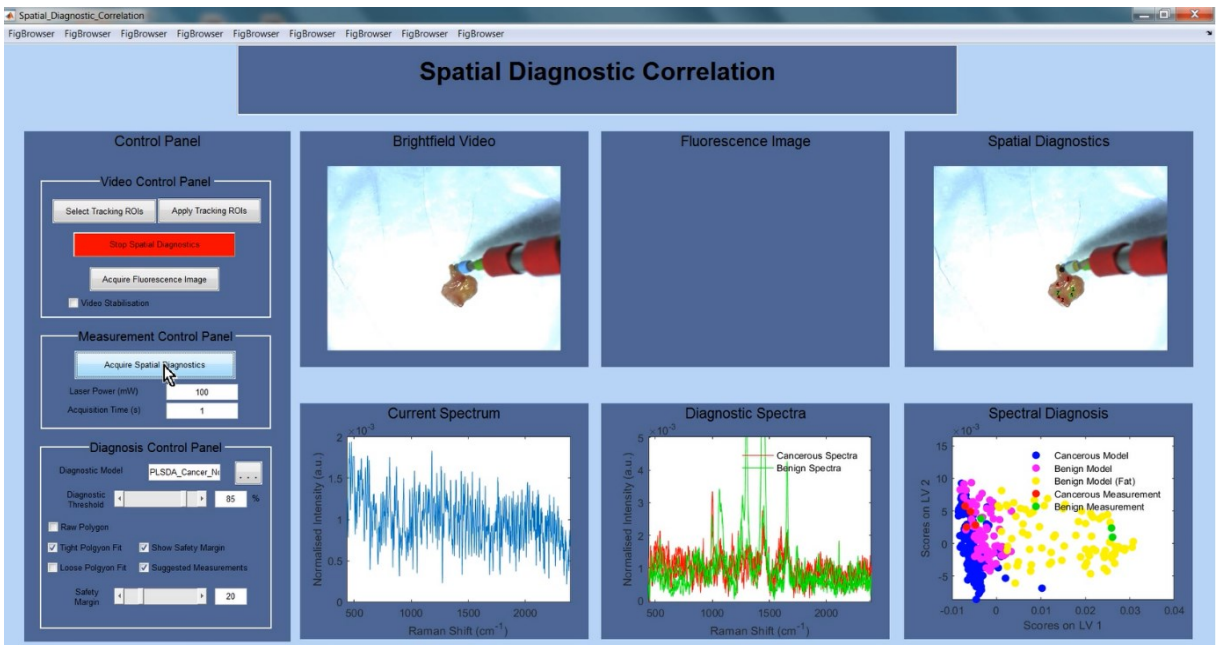

**Supplementary Video 1 | White Light-Guided Spectroscopic Margin Delineation *Ex Vivo*.** Video of white-light spectroscopic margin delineation of a human squamous cell carcinoma biopsy specimen.

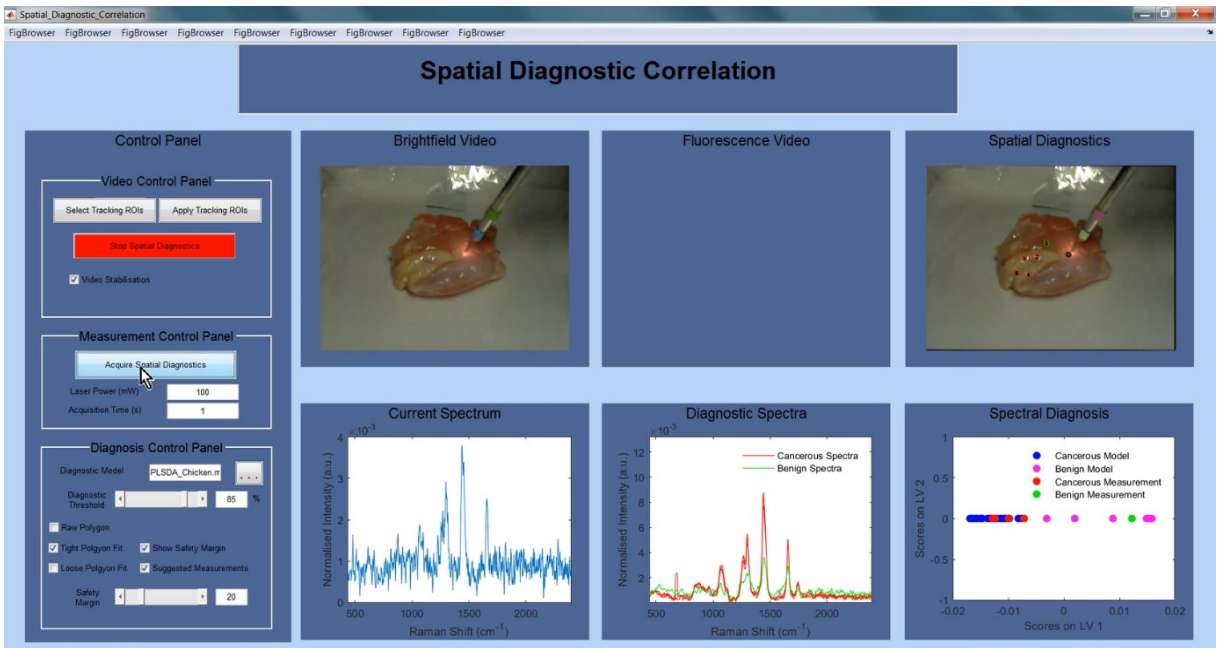

**Supplementary Video 2 | Demonstration of Fluorescence-Guided Raman Spectroscopic Diagnostic System Features.**
Video demonstrating implemented features for the fluorescence-guided Raman spectroscopic diagnostic system.